\documentclass{hotnets22}

\usepackage{titlesec}
\usepackage{microtype}  
\usepackage{amsmath,amsfonts}
\usepackage{algorithmic}
\usepackage{algorithm}
\usepackage{array}
\usepackage{textcomp}
\usepackage{stfloats}
\usepackage{url}
\usepackage{mathrsfs}  
\usepackage{verbatim}
\usepackage{graphicx}
\usepackage{hyperref}
\usepackage{mathtools}
\usepackage{tcolorbox}
\usepackage{color}
\usepackage{theorem}
\usepackage{times,epsfig}
\usepackage{amssymb}
\usepackage{cite}
\usepackage{cases}
\usepackage{todonotes}
\usepackage{xspace}
\usepackage{graphicx}  
\usepackage{subcaption}
\usepackage{tabularray}
\usepackage{tikz} 
\usepackage{enumitem}
\usepackage{todonotes}
\usepackage[font=bf]{caption}
\usepackage{csquotes}

\hypersetup{pdfstartview=FitH,pdfpagelayout=SinglePage}

\newcommand{\bm}{\textsf{NeMoEval}\xspace}

\newcommand*\circled[1]{\tikz[baseline=(char.base)]{
            \node[shape=circle,draw,inner sep=1.2pt] (char) {#1};}} 

\newcommand{\tightcaption}[1]{\vspace{-0.2cm}\caption{#1}\vspace{-0.2cm}}

\begin{document}


\title{Enhancing Network Management Using Code\\ Generated by Large Language Models}

\newcommand{\msft}{{\large$^\mathsection$}}
\newcommand{\bu}{{\large$^\dag$}}
\newcommand{\ru}{{\large$^\star$}}

\author{{Sathiya Kumaran Mani\msft \quad 
Yajie Zhou\msft\bu \quad 
Kevin Hsieh\msft
} \vspace{1mm}\\ 
{Santiago Segarra\msft\ru \quad
Ranveer Chandra\msft \quad 
Srikanth Kandula\msft}
\vspace{2mm} \\
{\it \msft Microsoft Research \quad \bu Boston University \quad \ru Rice University}}

\maketitle

\begin{abstract}
    Analyzing network topologies and communication graphs plays a crucial role in contemporary network management. However, the absence of a cohesive approach leads to a challenging learning curve, heightened errors, and inefficiencies. In this paper, we introduce a novel approach to facilitate a natural-language-based network management experience, utilizing large language models (LLMs) to generate task-specific code from natural language queries. This method tackles the challenges of explainability, scalability, and privacy by allowing network operators to inspect the generated code, eliminating the need to share network data with LLMs, and concentrating on application-specific requests combined with general program synthesis techniques. We design and evaluate a prototype system using benchmark applications, showcasing high accuracy, cost-effectiveness, and the potential for further enhancements using complementary program synthesis techniques. 
\end{abstract}

\section{Introduction}
\label{sec:intro}
A critical aspect of contemporary network management involves analyzing and performing actions on network topologies and communication graphs for tasks such as capacity planning~\cite{malt}, configuration analysis~\cite{batfish, minesweeper}, and traffic analysis~\cite{sonota, netseer,tpgs}. For instance, network operators may pose capacity planning questions, such as \enquote{What is the most cost-efficient way to double the network bandwidth between these two data centers?} using network topology data. Similarly, they may ask diagnostic questions like, \enquote{What is the required number of hops for data transmission between these two nodes?} using communication graphs. Network operators today rely on an expanding array of tools and domain-specific languages (DSLs) for these operations~\cite{malt,batfish}. A unified approach holds significant potential to reduce the learning curve and minimize errors and inefficiencies in manual operations.

The recent advancements in large language models (LLMs) \cite{gpt3, gpt4, palm, palm2, llama} provide a valuable opportunity to carry out network management tasks using natural language. LLMs have demonstrated exceptional proficiency in interpreting human language and providing high-quality answers across various domains \cite{gptaregpts, med-palm, codexdb, alphacode}. The capabilities of LLMs can potentially bridge the gap between diverse tools and DSLs, leading to a more cohesive and user-friendly approach to handling network-related questions and tasks.

Unfortunately, while numerous network management operations can be represented as graph analysis or manipulation tasks, no existing systems facilitate graph manipulation using natural language. Asking LLMs to directly manipulate network topologies introduces three fundamental challenges related to explainability, scalability, and privacy. First, explaining the output of LLMs and enabling them to reason about complex problems remain unsolved issues \cite{llm-survey}. Even state-of-the-art LLMs suffer from well-established problems such as hallucinations \cite{llm-hallcination} and making basic arithmetic mistakes \cite{math-word-problem, sparks-of-agi}. This complicates the process of determining the methods employed by LLMs in deriving answers and evaluating their correctness. Second, LLMs are constrained by limited token window sizes \cite{prompt-patterns}, which restrict their capacity to process extensive network topologies and communication graphs. For example, state-of-the-art LLMs such as Bard \cite{bard}, ChatGPT \cite{chatgpt}, and GPT-4 \cite{gpt4} permit only 2k to 32k tokens in their prompts, which can only accommodate small network topologies comprising tens of nodes and hundreds of edges. Third, network data may consist of personally identifiable information (PII), such as IP addresses \cite{gdpr}, raising privacy concerns when transferring this information to LLMs for processing. Addressing these challenges is crucial to develop a more effective approach to integrating LLMs in network management tasks.

\begin{figure*}[t]
    \centering   
    \begin{subfigure}{0.9\textwidth}  
        \centering  
        \includegraphics[width=\textwidth]{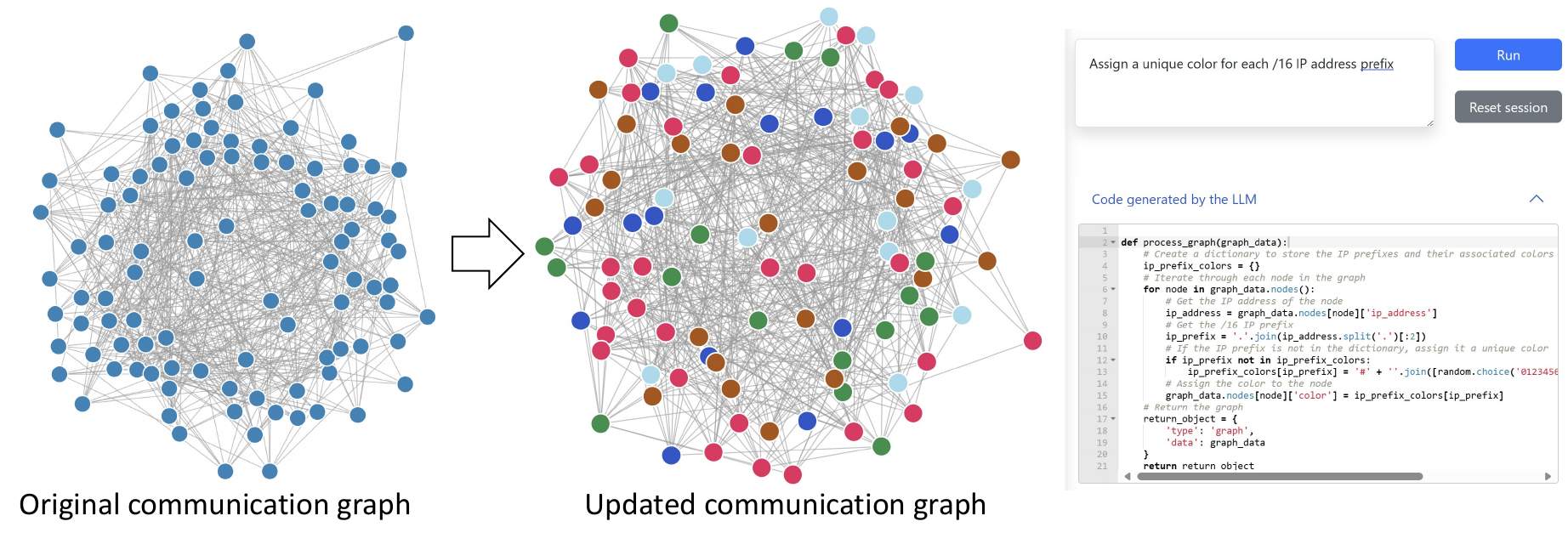}  
    \end{subfigure}  
    \caption{An example of how a natural-language-based network management system generates and executes a program in response to a network operator's query: \enquote{Assign a unique color for each /16 IP address prefix}. The system displays the LLM-generated code and the updated communication graph.}
    \label{fig:demo}
    \vspace{-0.3cm}
\end{figure*}

\noindent \textbf{Vision and Techniques}. In this paper, we present a novel approach to enhance network management by leveraging the power of LLMs to create \emph{task-specific code for graph analysis and manipulation}, which facilitates a natural-language-based network administration experience. Figure \ref{fig:demo} depicts an example of how this system generates and executes LLM-produced code in response to a network operator's natural language query. This approach tackles the explainability challenge by allowing network operators to examine the LLM-generated code, enabling them to comprehend the underlying logic and procedures to fulfill the natural language query. Additionally, it delegates computation to program execution engines, thereby minimizing arithmetic inaccuracies and LLM-induced hallucinations. Furthermore, this approach overcomes the scalability and privacy concerns by removing the necessity to transfer network data to LLMs, as the input for LLMs is the natural language query and the output solely comprises LLM-generated code.

The primary technical challenge in this approach lies in generating high-quality code that can reliably accomplish network management tasks. Although LLMs have demonstrated remarkable capabilities in general code generation \cite{alphacode,sparks-of-agi,mbpp}, they lack an understanding of domain-specific and application-specific requirements. To tackle this challenge, we propose a novel framework that combines application-specific requests with general program synthesis techniques to create customized code for graph manipulation tasks in network management. Our architecture divides the process of generating high-quality code into two key components: (1) an application-specific element that provides context, instructions, or plugins, which enhances the LLMs' comprehension of network structures, attributes, and terminology, and (2) a code generation element that leverages suitable libraries and cutting-edge program synthesis techniques \cite{human-eval, reflexion, self-consistency, self-debug, codet, mbpp} to produce code. This architecture fosters independent innovation of distinct components, and our preliminary study indicates substantial improvements in code quality.

\noindent \textbf{Implementation and Evaluation}. We design a prototype system that allows network operators to submit natural-language queries and obtain code to handle network topologies and communication graphs (Figure \ref{fig:demo}). To systematically assess effectiveness, we establish a benchmark, {\bm}, consisting of two applications that can be modeled as graph manipulation: (1) network traffic analysis using communication graphs \cite{tpgs, sonota, netseer}, and (2) network lifecycle management based on Multi-Abstraction-Layer Topology representation (MALT) \cite{malt}. To assess generalizability, we evaluate these applications using three code generation approaches (SQL \cite{sql}, pandas \cite{pandas}, and NetworkX \cite{networkx}) and four distinct LLMs \cite{human-eval, chatgpt, gpt4, bard}.
Our preliminary investigation shows that our system is capable of producing high-quality code for graph manipulation tasks. Utilizing the NetworkX-based approach, we attain average code correctness of 68\% and 56\% across all tasks for the four LLMs (up to 88\% and 78\% with GPT-4) for network traffic analysis and network lifecycle management, respectively.
In comparison, the strawman baseline, which inputs the limited-sized graph data directly into LLMs, only reaches an average correctness of 23\% for the traffic analysis application. Our method significantly improves the average correctness by 45\%, making it a more viable option.
Additionally, we demonstrate that integrating our system with complementary program synthesis methods could further enhance code quality for complex tasks.
Finally, we demonstrate that our approach is cost-effective, with an average expense of \$0.1 per task, and the LLM cost stays constant regardless of network sizes. Our study indicates that this is a potentially promising research direction. We release {\bm}\footnote{\label{nemo_eval_url}\url{https://github.com/microsoft/NeMoEval}}, our benchmark and datasets, to foster further research.

\noindent \textbf{Contributions}. We make the following contributions:

\begin{itemize}[topsep=1pt,itemsep=0ex,parsep=1ex, leftmargin=*]
    \item Towards enabling natural-language-based network administration experience, we introduce a novel approach that uses LLMs to generate code for graph manipulation tasks. This work is, to the best of our knowledge, the first to investigate the utilization of LLMs for graph manipulation and network management.
    \item We develop and release a benchmark that encompasses two network administration applications: network traffic analysis and network lifecycle management.
    \item We evaluate these applications with three code generation techniques and four distinct LLMs to validate the capability of our approach in generating high-quality code for graph manipulation tasks.
\end{itemize}

\section{Preliminaries}
\label{sec:background}

We examine graph analysis and manipulation's role in network management, followed by discussing recent LLM advances and their potential application to network management.

\subsection{Graph Analysis and Manipulation in Network Management}
\label{subsec:graph-manipulation}

Network management involves an array of tasks such as network planning, monitoring, configuration, and troubleshooting. As networks expand in size and complexity, these tasks become progressively more challenging. For instance, network operators are required to configure numerous network devices to enforce intricate policies and monitor these devices to guarantee their proper functionality. Numerous operations can be modeled as graph analysis and manipulation for network topologies or communication graphs. Two examples of these tasks are described below.

\noindent \textbf{Network Traffic Analysis.} Network operators analyze network traffic for various reasons, including identifying bottlenecks, congestion points, and underutilized resources, as well as performing traffic classification. A valuable representation in traffic analysis is traffic dispersion graphs (TDGs)\cite{tpgs} or communication graphs\cite{visualize-network-traffic}, in which nodes represent network components like routers, switches, or devices, and edges symbolize the connections or paths between these components (e.g., Figure \ref{fig:demo}). These graphs offer a visual representation of data packet paths, facilitating a comprehensive understanding of traffic patterns. Network operators typically utilize these graphs in two ways: (1) examining these graphs to understand the network's current state for network performance optimization\cite{tpgs}, traffic classification\cite{traffic-classification}, and anomaly detection\cite{tdgs-anomaly-detection}, and (2) manipulating the nodes and edges to simulate the impact of their actions on the network's performance and reliability\cite{network-monitoring}.

\noindent \textbf{Network Lifecycle Management.} Managing the entire lifecycle of a network entails various phases, including capacity planning, network topology design, deployment planning, and diagnostic operations. The majority of these operations necessitate an accurate representation of network topology at different abstraction levels and the manipulation of topology to achieve the desired network state~\cite{malt}. For example, network operators might employ a high-level topology to plan the network's capacity and explore different alternatives to increase bandwidth between two data centers. Similarly, network engineers may use a low-level topology to determine the location of a specific network device and its connections to other devices.

Hence, graph analysis and manipulation are crucial parts of network management. A unified interface capable of comprehending and executing these tasks has the potential to significantly simplify the process, saving network operators considerable time and effort.

\subsection{LLMs and Program Synthesis}
\label{subsec:program-synthesis}

Automated program generation based on natural language descriptions, also known as program synthesis, has been a long-standing research challenge \cite{fortran-automatic-coding-system, automatic-program-synthesis, program-synthesis}. Until recently, program synthesis had primarily been limited to specific domains, such as string processing \cite{automatic-string-processing}, program generation based on input-output examples \cite{deepcoder}, and natural language for database queries (e.g., \cite{interactive-nl-interface, learning-sql, nl2sql-where-are-we}). In contrast, general program synthesis was considered to be out of reach \cite{mbpp}. The breakthrough emerged with the advancement of LLMs \cite{human-eval,gpt3,gpt4,incoder,starcoder,bard}, which are trained on extensive corpora of natural language text from the internet and massive code repositories such as GitHub. LLMs have demonstrated remarkable proficiency in learning the relationship between natural language and code, achieving state-of-the-art performance in domain-specific tasks such as natural language to database query \cite{sql-palm, lever}, as well as human-level performance in tasks like programming competitions \cite{alphacode} and mock technical interviews \cite{sparks-of-agi}. Just recently, these advancements have led to experimental plugins designed to solve mathematical problems and perform data analysis through code generation \cite{code-interpreter}.

The recent breakthrough in program synthesis using LLMs has ignited a surge of research aimed at advancing the state of the art in this field. These techniques can generally be classified into three approaches: (1) code selection, which involves generating multiple samples with LLMs and choosing the best one based on the consistency of execution results~\cite{self-consistency} or auto-generated test cases~\cite{codet}; (2) few-shot examples, which supply LLMs with several examples of the target program's input-output behavior~\cite{mbpp}; and (3) feedback and self-reflection, which incorporates a feedback or reinforcement learning outer loop to help LLMs learn from their errors~\cite{self-debug, reflexion, code-gen-with-feedback}. These advanced techniques continue to expand the horizons of program synthesis, empowering LLMs to generate more complex and accurate programs. 

As Section \ref{sec:intro} discusses, LLM-generated code can tackle explainability, scalability, and privacy challenges in LLM-based network management. However, our initial study shows that merely applying existing approaches is inadequate for network management tasks, as existing techniques do not comprehend the domain-specific and application-specific requirements. The key technical challenge lies in harnessing recent advancements in LLMs and general program synthesis to develop a unified interface capable of accomplishing network management tasks, which forms the design requirements for our proposed solution.
\begin{figure*}[t]
    \centering  
    \includegraphics[width=0.8\textwidth]{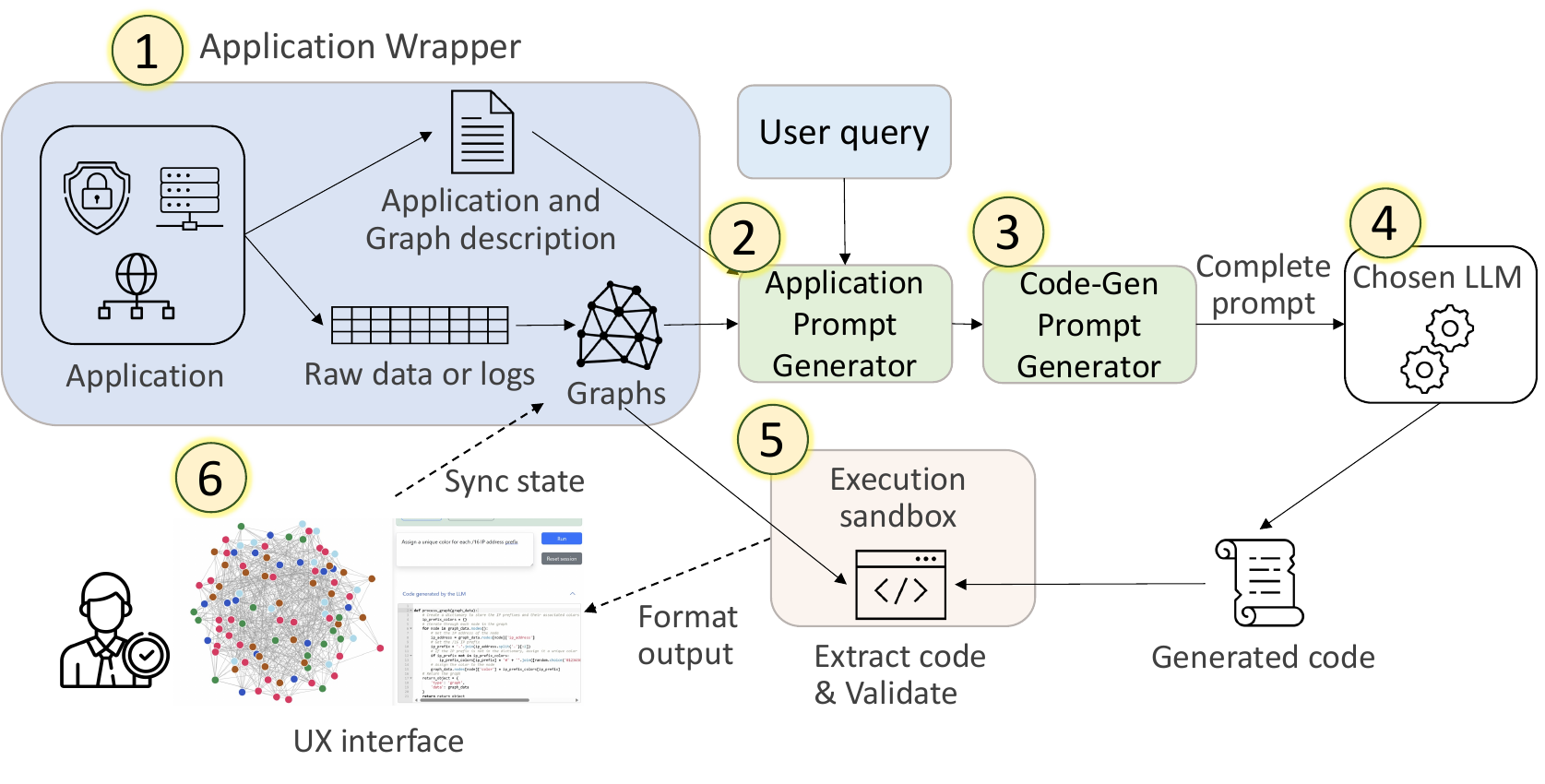}  
    \caption{A general framework for network management systems using natural language and LLM-generated code}
    \label{fig:solution}
\end{figure*}

\section{System Framework}

We present a novel system framework designed to enhance network management by utilizing LLMs to generate task-specific code. Our framework is founded on two insights. First, we can transform many network management operations into graph analysis and manipulation tasks (Section \ref{subsec:graph-manipulation}), which allows for a unified design and a more focused task for code generation. Second, we can divide prompt generation into two aspects: domain-specific requirements and general program synthesis. By combining the strengths of domain specialization with recent advances in program synthesis techniques (Section \ref{subsec:program-synthesis}), we can generate high-quality code for network management tasks. Figure \ref{fig:solution} illustrates our system framework.

The framework we propose consists of an application wrapper (\circled{1} in Figure \ref{fig:solution}) that uses domain-specific knowledge, such as the definitions of nodes and edges, to transform the application data into a graph representation. This information, together with user queries in natural language, is processed by an application prompt generator (\circled{2}) to create a task-specific prompt for the LLM. Subsequently, the task-specific prompt is combined with a general code-gen prompt generator (\circled{3}) to instruct the LLM (\circled{4}) to produce code. The generated code utilizes plugins and libraries to respond to the user's natural language queries in the constructed graph. An execution sandbox (\circled{5}) executes the code on the graph representation of the network. The code and its results are displayed on a UX interface (\circled{6}). If the user approves the results, the UX sends the updated graph back to the application wrapper (\circled{1}) to modify the network state and record the input/output for future prompt enhancements~\cite{mbpp,self-debug,reflexion}. We describe the key components below.

\noindent \textbf{Application Wrapper (\circled{1}).} The application wrapper offers context-specific information related to the network management application and the network itself. For instance, the Multi-Abstraction-Layer Topology representation (MALT) wrapper\cite{malt} can extract the graph of entities and relationships from the underlying data, describing entities (e.g., packet switches, control points, etc.) and relationships (e.g., contains, controls, etc.) in natural language. This information assists LLMs in comprehending the network management application and the graph data structure. Additionally, the application wrapper can provide application-specific plugins~\cite{chatgpt-plugin} or code libraries to make LLM tasks more straightforward.

\noindent \textbf{Application Prompt Generator (\circled{2}).} The purpose of the application prompt generator is to accept both the user query and the information from the application wrapper as input, and then generate a prompt specifically tailored to the query and task for the LLM. To achieve this, the prompt generator can utilize a range of static and dynamic techniques\cite{guidance, chain-of-thought, auto-chain-of-thought}. For instance, when working with MALT, the prompt generator can dynamically select relevant entities and relationships based on the user query, and then populate a prompt template with the contextual information. Our framework is designed to offer flexibility regarding the code-gen prompt generator (\circled{3}) and LLMs (\circled{4}), enabling the use of various techniques for different applications.

\noindent \textbf{Execution Sandbox (\circled{5}).} As highlighted in previous research~\cite{human-eval}, it is crucial to have a secure environment to run the code generated by LLMs. The execution sandbox can be established using virtualization or containerization techniques, ensuring limited access to program libraries and system calls. Additionally, this module provides a chance to enhance the security of both code and system by validating network invariants or examining output formats.

\section{Implementation and Evaluation}
\label{sec:eval}

\subsection{Benchmark}
\label{subsec:benchmark}

We design a general benchmark, {\bm}, to evaluate the effectiveness of LLM-based network management systems. Figure \ref{fig:benchmark} illustrates the architecture of our benchmark, which consists of three primary components:
\begin{figure}[h]
    \centering  
    \begin{subfigure}{0.9\columnwidth}  
        \centering  
        \includegraphics[width=\textwidth]{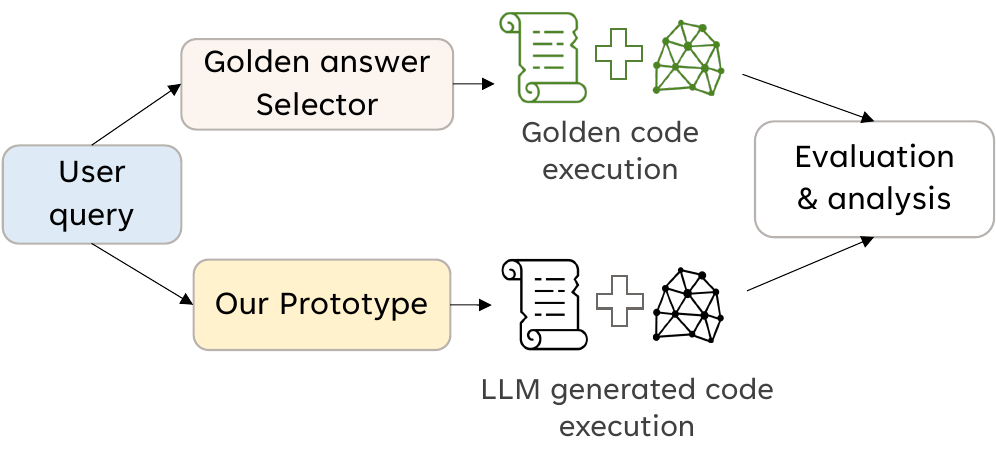}  
    \end{subfigure}
    \vspace{-0.2cm}
    \tightcaption{Benchmark design}
\label{fig:benchmark}  
\end{figure}

\noindent\textbf{Golden Answer Selector.} For each input user query, we create a \enquote{golden answer} with the help of human experts. These verified answers, stored in a selector's dictionary file, act as the ground truth to evaluate LLM-generated code.

\noindent\textbf{Results Evaluator.} The system executes the LLM-generated code on network data, comparing outcomes with the golden answer's results. If they match, the LLM passes; otherwise, it fails, and we document the findings for further analysis.

\noindent\textbf{Results Logger.} To facilitate the analysis of the LLM's performance and the identification of potential improvement, we log the results of each query, including the LLM-generated code, the golden answer, and the comparison results. The results logger also records any code execution errors that may have occurred during the evaluation process. 

\subsection{Experimental Setup}
\label{subsec:setup}

\noindent\textbf{Applications and Queries.} We implement and evaluate two applications, network traffic analysis and network lifecycle management (Section \ref{subsec:graph-manipulation}):

\begin{itemize}[topsep=1pt,itemsep=0ex,parsep=1ex, leftmargin=*]
    \item \emph{Network Traffic Analysis.} We generate synthetic communication graphs with varying numbers of nodes and edges. Each edge represents communication activities between two nodes with random weights in bytes, connections, and packets. We develop 24 queries by curating trial users' queries, encompassing common tasks such as topology analysis, information computation, and graph manipulation.
    \item \emph{Network Lifecycle Management.} We use the example MALT dataset~\cite{malt-data-link} and convert it into a directed graph with 5493 nodes and 6424 edges. Each node represents one or more types in a network, such as packet switches, chassis, and ports, with different node types containing various attributes. Directed edges encapsulate relationships between devices, like control or containment associations. We develop 9 network management queries focusing on operational management, WAN capacity planning, and topology design.
\end{itemize}

The queries are categorized into three complexity levels (\enquote{Easy}, \enquote{Medium}, and \enquote{Hard}) based on the complexity of their respective golden answers. Table \ref{tab::prompt_list} displays an example query from each category due to page limits. We release the complete list of queries, their respective golden answers, and the benchmark to facilitate future research\footnote{\url{https://github.com/microsoft/NeMoEval}}.
\begin{table*}
\centering
\caption{User query examples. See all queries in {\bm}.}
\resizebox{1\textwidth}{!}{
\begin{tblr}{
  cells = {c},
  hline{1-2,5} = {-}{},
}
Complexity level & Traffic Analysis                                                                                                         & MALT                                                                                          \\
Easy             & Add a label app:prodution to nodes with address prefix 15.76 & List all ports that are contained by packet switch ju1.a1.m1.s2c1.                            \\
Medium           & Assign a unique color for each /16 IP address prefix.                                                                    & Find the first and the second largest Chassis by capacity.                                    \\
Hard             & Calculate total byte weight on each node, cluster them into 5 groups.                          & Remove packet switch P1 from Chassis 4, balance the capacity afterward. 
\end{tblr}
}
\label{tab::prompt_list}
\end{table*}
\begin{table}[t]
\centering
\tightcaption{Accuracy Summary for Both Applications}
\resizebox{1\columnwidth}{!}{
\begin{tblr}{
  column{even} = {c},
  column{3} = {c},
  column{5} = {c},
  column{7} = {c},
  cell{1}{2} = {c=4}{},
  cell{1}{6} = {c=3}{},
  hline{1,3,7} = {-}{},
}
                 & Traffic Analysis &      &        &               & MALT &               &               \\
                 & Strawman         & SQL  & Pandas & NetworkX      & SQL  & Pandas        & NetworkX      \\
GPT-4            & 0.29             & 0.50 & 0.38   & \textbf{0.88} & 0.11 & 0.56          & \textbf{0.78} \\
GPT-3            & 0.17             & 0.13 & 0.25   & \textbf{0.63} & 0.11 & \textbf{0.44} & \textbf{0.44} \\
text-davinci-003 & 0.21             & 0.29 & 0.29   & \textbf{0.63} & 0.11 & 0.22          & \textbf{0.56} \\
Google Bard        & 0.25             & 0.21 & 0.25   & \textbf{0.59} & 0.11  & 0.33          & \textbf{0.44} 
\end{tblr}
}
\label{tab::sum_together}
\end{table}
\begin{table}[t]
\centering
\tightcaption{Breakdown for Trafic Analysis}
\resizebox{1\columnwidth}{!}{
\begin{tblr}{
  column{even} = {c},
  column{3} = {c},
  column{5} = {c},
  hline{1,3,7} = {-}{},
}
                 & Strawman       & SQL            & Pandas         & NetworkX       \\
                 & E(8)/M(8)/H(8) & E(8)/M(8)/H(8) & E(8)/M(8)/H(8) & E(8)/M(8)/H(8) \\
GPT-4            & 0.50/0.38/0.0  & 0.75/0.50/0.25 & 0.50/0.50/0.13 & 1.0/1.0/0.63   \\
GPT-3            & 0.38/0.13/0.0  & 0.25/0.13/0.0  & 0.50/0.25/0.0  & 1.0/0.63/0.25  \\
text-davinci-003 & 0.38/0.25/0.0  & 0.63/0.25/0.0  & 0.63/0.25/0.0  & 1.0/0.75/0.13  \\
Google Bard      & 0.50/0.25/0.0  & 0.38/0.25/0.0  & 0.50/0.13/0.13 & 0.88/0.50/0.38 
\end{tblr}
}
\label{tab::break_traffic}
\end{table}
\begin{table}[t]
  \centering
  \tightcaption{Breakdown for MALT}
  \resizebox{1\columnwidth}{!}{
  \begin{tblr}{
    column{even} = {c},
    column{3} = {c},
    hline{1,3,7} = {-}{},
  }
                   & SQL            & Pandas         & NetworkX       \\
                   & E(3)/M(3)/H(3) & E(3)/M(3)/H(3) & E(3)/M(3)/H(3) \\
  GPT-4            & 0.33/0.0/0.0   & 0.67/0.67/0.33 & 1.0/1.0/0.33   \\
  GPT-3            & 0.33/0.0/0.0   & 0.67/0.67/0.0  & 0.67/0.67/0.0  \\
  text-davinci-003 & 0.33/0.0/0.0   & 0.33/0.33/0.0  & 0.67/0.67/0.33 \\
  Google Bard      & 0.33/0.0/0.0   & 0.67/0.33/0.0  & 0.67/0.33/0.33  
  \end{tblr}
  }
  \label{tab::break_malt}
  \end{table}

\noindent \textbf{LLMs.} We conduct our study on four state-of-the-art LLMs, including GPT-4 \cite{gpt4}, GPT-3 \cite{gpt3}, Text-davinci-003 (a variant of GPT 3.5) \cite{openai-models}, and Google Bard \cite{bard}. We further explore two open LLMs, StarCoder \cite{starcoder} and InCoder \cite{incoder}. However, we do not show their results here because of inconsistent answers. We intend to report their results once they achieve consistent performance in future investigation. With all OpenAI LLMs, we set their temperature to 0 to ensure consistent output across multiple trials. Since we cannot change the temperature of Google Bard, we send each query five times and calculate the average passing probability \cite{human-eval}.

\noindent \textbf{Approaches.} We implement three code generation methods using well-established data/graph manipulation libraries, which offer abundant examples in public code repositories for LLMs to learn from:

\begin{itemize}[topsep=1pt,itemsep=0ex,parsep=1ex, leftmargin=*]
    \item \emph{NetworkX.} We represent the network data as a NetworkX~\cite{networkx} graph, which offers flexible APIs for efficient manipulation and analysis of network graphs.
    \item \emph{pandas.} We represent the network data using two pandas~\cite{pandas} dataframes: a node dataframe, which stores node indices and attributes, and an edge dataframe, which encapsulates the link information among nodes through an edge list. Pandas provides many built-in data manipulation techniques, such as filtering, sorting, and grouping.
    \item \emph{SQL.} We represent the network data as a relational database queried through SQL~\cite{sql}, consisting of a table for nodes and another for edges. The table schemas are similar to those in pandas. Recent work has demonstrated that LLMs are capable of generating SQL queries with state-of-the-art accuracy \cite{lever,sql-palm}.
\end{itemize}

We also evaluate an alternative baseline (\emph{strawman}) that directly feeds the original network graph data in JSON format to the LLM and requests it to address the query. However, owing to the token constraints on LLMs, we limit our evaluation of this approach to synthetic graphs for network traffic analysis, where data size can be controlled.

\subsection{Code Quality}

Table \ref{tab::sum_together} summarizes the benchmark results for network traffic analysis and network lifecycle management, respectively. We observe three key points. First, utilizing LLMs for generating code in network management significantly surpasses the strawman baseline in both applications, as the generated code reduces arithmetic errors and LLM hallucinations. Second, employing a graph library (NetworkX) greatly enhances code accuracy compared to pandas and SQL, as LLMs can leverage NetworkX's graph manipulation APIs to simplify the generated code. This trend is consistent across all four LLMs. Finally, pairing NetworkX with the state-of-the-art GPT-4 model produces the highest results (88\% and 78\%, respectively), making it a promising strategy for network management code generation.

To understand the impact of task difficulty, we break down the accuracy results in Tables \ref{tab::break_traffic} and \ref{tab::break_malt}. We observe that the accuracy of LLM-generated code decreases as task complexity increases. This trend is consistent across all LLMs and approaches, with the performance disparities becoming more pronounced for network lifecycle management (Table \ref{tab::break_malt}). 

Our analysis of the LLM-generated code reveals that the complex relationships in the MALT dataset make LLMs more prone to errors in challenging tasks, and future research should focus on improving LLMs' ability to handle complex network management tasks.

\begin{table}
\caption{Error Type Summary of LLM Generated Code}
\resizebox{1\columnwidth}{!}{
\begin{tblr}{
  hline{1-2,9} = {-}{},
}
LLM's error type (NetworkX)                                           & Traffic Analysis (35)    & MALT (17) \\
Syntax error                & 9  & 0    \\
Imaginary graph attributes          & 9                   & 1    \\
Imaginary files/function arguments        & 3        & 2    \\
Arguments error       & 7                   & 8    \\
Operation error & 4                   & 2    \\
Wrong calculation logic                                    & 2                   & 3   \\
Graphs are not identical                                   & 1                   & 1    \\    
\end{tblr}
}
\label{tab::error_type}
\vspace{0.2cm}
\end{table}
\begin{table}
\centering
\tightcaption{Improvement Cases with Bard on MALT}
\resizebox{1\columnwidth}{!}{
\begin{tabular}{cccc} 
\hline
         & Bard + Pass@1 & Bard + Pass@5 & Bard + Self-debug  \\
NetworkX & 0.44        & 1.0          & 0.67               \\
\hline
\end{tabular}
}
\label{tab::sum_improve}
\vspace{-0.3cm}
\end{table}

\subsection{Case Study on Potential Improvement}

For the NetworkX approach across all four LLMs, there are 35 failures out of 96 tests ($24\times4$) for network traffic analysis and 17 failures out of 36 tests ($9\times4$) for network lifecycle management, respectively.
Table \ref{tab::error_type} summarizes the error types.
More than half of the errors are associated with syntax errors or imaginary (non-existent) attributes. We conduct a case study to see whether using complementary program synthesis techniques (Section \ref{subsec:program-synthesis}) could correct these errors. 

We assess two techniques: (1) pass@k \cite{human-eval}, where the LLM is queried $k$ times with the same question, and it is deemed successful if at least one of the answers is correct. This method reduces errors arising from the LLM's inherent randomness and can be combined with code selection techniques \cite{human-eval,codet,self-consistency} for improved results; (2) self-debug \cite{self-debug}, which involves providing the error message back to the LLM and encouraging it to correct the previous response.

We carry out a case study using the Bard model and three unsuccessful network lifecycle queries with the NetworkX approach. Table \ref{tab::sum_improve} shows that both pass@k ($k=5$) and self-debug significantly enhance code quality, resulting in improvements of 100\% and 67\%, respectively. These results indicate that applying complementary techniques has considerable potential for further improving the accuracy of LLM-generated code in network management applications.

\subsection{Cost and Scalability Analysis}

\begin{figure}
    \centering
    \begin{subfigure}{0.7\columnwidth}
    {
        \includegraphics[width=0.99\linewidth]{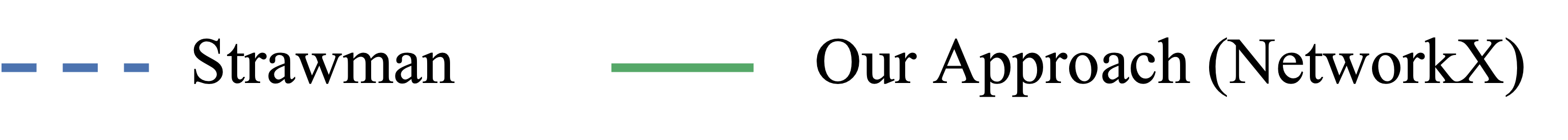}
        \vspace{-15pt}
    }
    \end{subfigure}
    
    \begin{subfigure}[t]{0.45\columnwidth}
    {
        \includegraphics[width=1\linewidth]{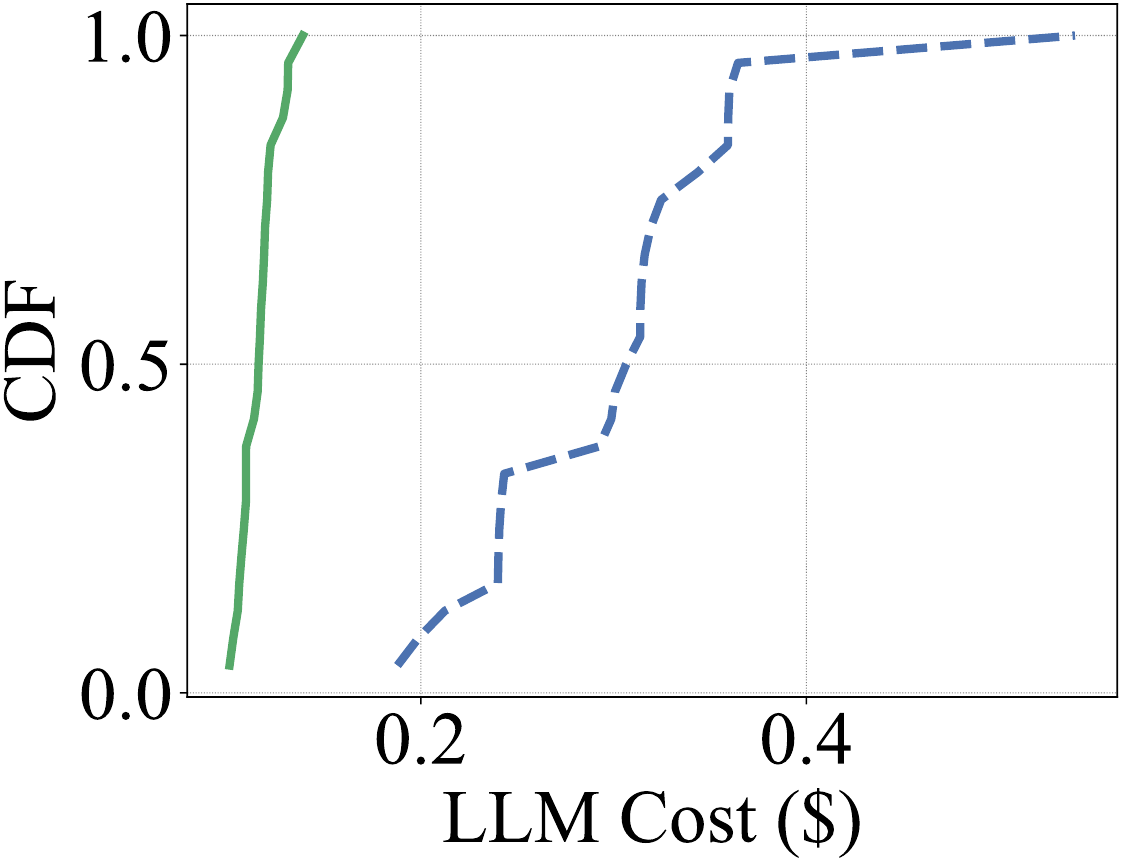}
        \caption{CDF of LLM cost per query (80 nodes and edges)}
        \label{fig::cost_analysis_a}

    }
    \end{subfigure}
    \smallskip
    \begin{subfigure}[t]{0.45\columnwidth}
    {
        \includegraphics[width=1\linewidth]{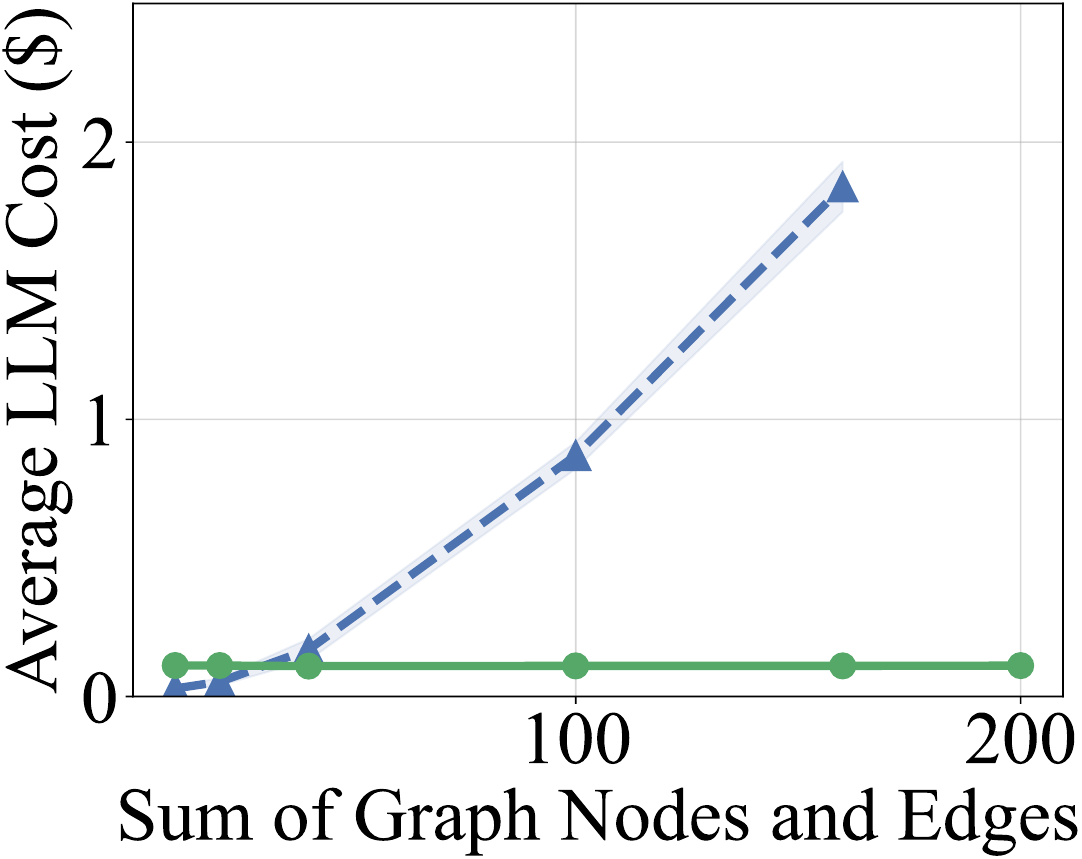}
        \caption{Cost analysis on graph size}
        \label{fig::cost_analysis_b}

    }
    \end{subfigure}
    \tightcaption{Cost and scalability Analysis}
    \label{fig::cost_analysis}
    \vspace{-1mm}
\end{figure}

We examine the LLM cost utilizing GPT-4 pricing on Azure \cite{openai-pricing} for the network traffic analysis application. Figure \ref{fig::cost_analysis_a} reveals that the strawman approach is three times costlier than our method for a small graph with 80 nodes and edges. As the graph size expands (Figure \ref{fig::cost_analysis_b}), the gap between the two approaches grows, with the strawman approach surpassing the LLM's token limit for a moderate graph containing 150 nodes and edges. Conversely, our method has a small cost (<\$0.2 per query) that remains unaffected by graph size increases.

\section{Discussion and Conclusion}

Recent advancement in LLMs has paved the way for new opportunities in network management. We introduce a system framework that leverages LLMs to create task-specific code for graph manipulation, tackling issues of explainability, scalability, and privacy. While our prototype and preliminary study indicate the potential of this method, many open questions remain in this nascent area of research.

\noindent \textbf{Code Quality for Complex Tasks.} As our evaluation demonstrates, the LLM-generated code is highly accurate for easy and medium tasks; however, the accuracy decreases for more complex tasks. This is partially due to the LLMs being trained on a general code corpus without specific network management knowledge. An open question is how to develop domain-specific program synthesis techniques capable of generating high-quality code for complex network management tasks, such as decomposing the task into simpler sub-tasks\cite{chain-of-thought}, incorporating application-specific plugins\cite{chatgpt-plugin}, or fine-tuning the model with application-specific code examples.

\noindent \textbf{Code Comprehension and Validation.} Ensuring correctness and understanding LLM-generated code can be challenging for network operators. While general approaches like LLM-generated test cases\cite{codet} and code explanation~\cite{github-copilot-x} exist, they are insufficient for complex tasks. Developing robust, application-specific methods to aid comprehension and validation is a crucial challenge.

\noindent \textbf{Expanding Benchmarks and Applications.} Extending our current benchmark to cover more network management tasks raises questions about broader effectiveness and applicability to other applications, such as network failure diagnosis~\cite{passive-fault-detection, shrink} and configuration verification~\cite{minesweeper,batfish}. Addressing these challenges requires exploring new network state representation, code generation strategies, and application-specific libraries and plugins.

In summary, we take a pioneering step in introducing a general framework to use LLMs in network management, presenting a new frontier for simplifying network operators' tasks. We hope that our work, along with our benchmarks and datasets, will stimulate continued exploration in this field.

\bibliographystyle{abbrv} 
\begin{small}
\bibliography{ztncopilot}
\end{small}

\end{document}